\documentclass{jfm}
\usepackage{amsmath}
\newcommand*{\pd}[3][]{\ensuremath{\frac{\partial^{#1} #2}{\partial #3}}}
\usepackage{epstopdf}
\begin{document}

\newtheorem{lemma}{Lemma}
\newtheorem{corollary}{Corollary}

\shorttitle{On the Maxey--Riley equation of motion} 
\shortauthor{A. Talaei and T. Garrett} 

\title{On the Maxey--Riley equation of motion and its extension to high Reynolds numbers}

\author
 {
 Ahmad Talaei
  \corresp{\email{ahmad.talaei@utah.edu}},
  \and
  Timothy J. Garrett
  }

\affiliation
{
Department of Atmospheric Sciences, University of Utah, UT, USA
}

\maketitle

\begin{abstract}
The inertial response of a particle to turbulent flows is a problem of relevance to a wide range of environmental and engineering problems. The equation most often used to describe the force balance is the Maxey--Riley equation, which includes in addition to buoyancy and steady drag forces, an unsteady Basset drag force related to past particle acceleration. Here we provide a historical review of how the Maxey--Riley equation was developed and how it is only suited for studies where the Reynolds number is less than unity. Revisiting the innovative mathematical methods employed by Basset (1888), we introduce an alternative formulation for the unsteady drag for application to a broader range of particle motions. While the Basset unsteady drag is negligible at higher Reynolds numbers, the revised unsteady drag is not.
\end{abstract}

\section{Introduction}
Determination of a particle's trajectory in a turbulent flow field requires an equation that satisfies the Navier--Stokes equation and accounts for all relevant forces. The first attempt was made by Stokes for a sphere moving slowly with a uniform velocity in a viscous fluid of unlimited extent that is stationary far from the particle \citep{stokes1850effect}. Boussinesq and Basset later considered the linear inertia of flow surrounding the sphere and developed an equation for the unsteady motion of a spherical particle accelerating from rest and moving with a time-varying velocity $v_p(t)$, adding an unsteady drag force or ``history term" to the equation of motion that accounts for prior particle interactions with the surrounding flow \citep{boussinesq1885applications,boussinesq1885resistance,basset1888treatise}.

In the interests of mathematical simplicity, the derivation by \cite{boussinesq1885applications,boussinesq1885resistance} and \cite{basset1888treatise} omitted non-linear inertia terms proportional to the squares and products of velocities of the surrounding flow relative to a moving sphere. Such an assumption can be valid in the Stokes flow regime because the particle motion can be considered to be ``slow". Fluid viscous forces dominate inertia and the Reynolds number is ``small", i.e, $\Rey ={v_p d_p}/{\nu} < 1$ where $d_p$ is the sphere diameter and $\nu = \mu/\rho_f$ is the kinematic viscosity of the fluid, $\mu$ the dynamic viscosity of the fluid, and $\rho_f$ the fluid density.

The next significant advance was introduced by \cite{Tchen1947mean} who generalized the equation of motion for unsteady motion of spherical particle in a fluid at rest. He proposed an equation for the motion of a slow spherical particle in a fluid that has a velocity $v_f(t)$ independent of  the sphere. To reduce the problem to that of a particle moving in a fluid at rest, Tchen assumed the particle moves with a velocity $v_p(t)-v_f(t)$. In addition, he allowed for the entire system, including both the fluid and the particle to experience a pressure gradient force due to a changing rectilinear velocity of the fluid $v_f(t)$. \cite{corrsin1956equation} later showed that if the fluid is turbulent, and the sphere is smaller than the shortest wavelength characterizing the turbulent flow, spatial and temporal inhomogeneities in the fluid also add a torque due to spatial velocity gradients, and a force due to a static pressure gradient.

Further adaptations and extensions of the equation of motion account for the drag force due to the forced velocity curvature around the sphere, or the Faxén correction, and viscous shear stress, leading to the widely used Maxey--Riley equation \citep{faxen1922widerstand,buevich1966motion,Riley1971,soo1975equation,gitterman1980memory,maxey1983equation}. For a particle that is at rest in a stationary fluid until the instant $t=0$, and is sufficiently small to have a negligible effect on fluid motions far from the particle, the Maxey--Riley equation accounts for the trajectory, dispersion, and settling velocity of the particle. The force balance includes the buoyancy force, the stress gradient of the fluid flow in the absence of a particle, the force due to the virtual mass, steady Stokes drag and unsteady Basset drag

\begin{align}
    m_p\frac{d\textit{\textbf{v}}_p}{dt} =& (m_p-\rho_f V_p)\textit{\textbf{g}} + \rho_f V_p\frac{D\textit{\textbf{v}}_f}{Dt} - k\rho_f V_p\frac{d}{dt}\Big(\textit{\textbf{v}}_p-\textit{\textbf{v}}_f -\frac{1}{10} {a}^2 \nabla^2\textit{\textbf{v}}_f \Big) 
    \nonumber \\
    & -6 \pi \mu a \Big( \textit{\textbf{v}}_p-\textit{\textbf{v}}_f - \frac{1}{6} {a}^2 \nabla^2\textit{\textbf{v}}_f \Big) -6 \pi \mu a^2 \int_{0}^{t}\frac{\frac{d}{d\tau}(\textit{\textbf{v}}_p(\tau)-\textit{\textbf{v}}_f(\tau)-\frac{1}{6} {a}^2 \nabla^2\textit{\textbf{v}}_f)}{\sqrt{\pi \nu(t-\tau)}} d\tau
    \label{eq:Maxey--Riley}
\end{align}
where the index \textit{p} denotes the particle and \textit{f} for the fluid. $\frac{d}{dt}=\pd{}{t}+{v_p}_j \pd{}{x_j}$ is the total time derivative along the particle trajectory, and $\frac{D}{Dt}=\pd{}{t}+{v_f}_j \pd{}{x_j}$ the acceleration of the fluid along its own trajectory, $m_p$ the particle mass, $v_p$ the Lagrangian velocity of the particle, and $v_f$ the Eulerian fluid velocity in the particle location. $\rho_{f}V_p$ is the fluid mass $m_f$ occupied by the particle volume $V_p$ of radius $a$. $k=(m'_f/m_f)$ is an added mass coefficient, and $m'_f$ the virtual mass of the fluid, assumed to undergo the same acceleration as the particle. The coefficient $k$ is a function of the flow regime and geometric properties of the particle. For irrotational flow around a sphere $k$ is $0.5$.

No analytical solution exists for the full expression of the Maxey--Riley equation of motion. Numerically, however, the equation provides a useful guide for exploring interactions between particles and a moving fluid flow. Its application extends to fields as wide ranging as sediment transport and waste management, combustion, particle transport, and deposition, particle clustering, atmospheric precipitation, aquatic organism behaviors, and underwater robotics \citep{chao1963turbulent,soo1975equation,murray1970settling,reeks1977dispersion,nir1979effect,kubie1980settling,maxey1987gravitational,maxey1990advection,mei1990particle,mei1991particle,falkovich2002acceleration,peng2009transport,daitche2013advection,beron2019building}.

An important point is that \eqref{eq:Maxey--Riley} assumes that the Reynolds number of the particle relative to the surrounding fluid flow satisfies $\Rey = |\textit{\textbf{v}}_p-\textit{\textbf{v}}_f| d_p/\nu < 1$, that is the Stokes flow regime. For larger values of $\Rey$, semi-empirical adjustments are sometimes made \citep{ho1964fall,hwang1985fall,tunstall1968retardation,field1968effects,murray1970settling,maxey1990advection,wang1993settling,nielsen1993turbulence,stout1995effect,good2014settling}. The steady drag force $F_d = 6\pi\mu a(v_p - v_f)$ shifts from scaling linearly with relative velocity to scaling as an empirically derived steady drag coefficient $C_D$ and the relative velocity squared $F_d = \frac{1}{2} \rho_f A_p C_D(\Rey) (v_p - v_f)^2$. Then the steady drag at high Reynolds numbers becomes sufficiently large that the history term in \eqref{eq:Maxey--Riley} becomes negligible and can be omitted from the equation of motion \citep{wang1993settling,stout1995effect,good2014settling}. 

However, it remains that the mathematical form of the history term developed by Boussinesq-Basset applies only when the Reynolds number is small $(\mathrm{Re} < 1)$. A priori, there is no mathematical justification for arguing that unsteady drag is negligible compared to steady drag when the Reynolds number is arbitrarily high. While a few theoretical studies have considered unsteady drag on a sphere at a finite but small Reynolds number in the range $\mathrm{Re} < 100$ \citep{oseen1913ueber,proudman1957expansions,sano1981unsteady,mei1991unsteady,mei1992flow,mei1994flow,lovalenti1993force,michaelides1997transient}, as of yet, no general formulation has been presented for the unsteady drag on solid bodies moving within a viscous liquid when $\mathrm{Re} \gg 1$. This article attempts to fill this gap by first revisiting the classical derivation of Basset's solution, and then by using a similar approach obtaining a formulation for the unsteady drag term suitable for application to higher Reynolds numbers.

\section{Overview of the Stokes solution} \label{sec:Stokes-solution}
\begin{figure}
    \centering
    \includegraphics[width=6cm, height=7cm]{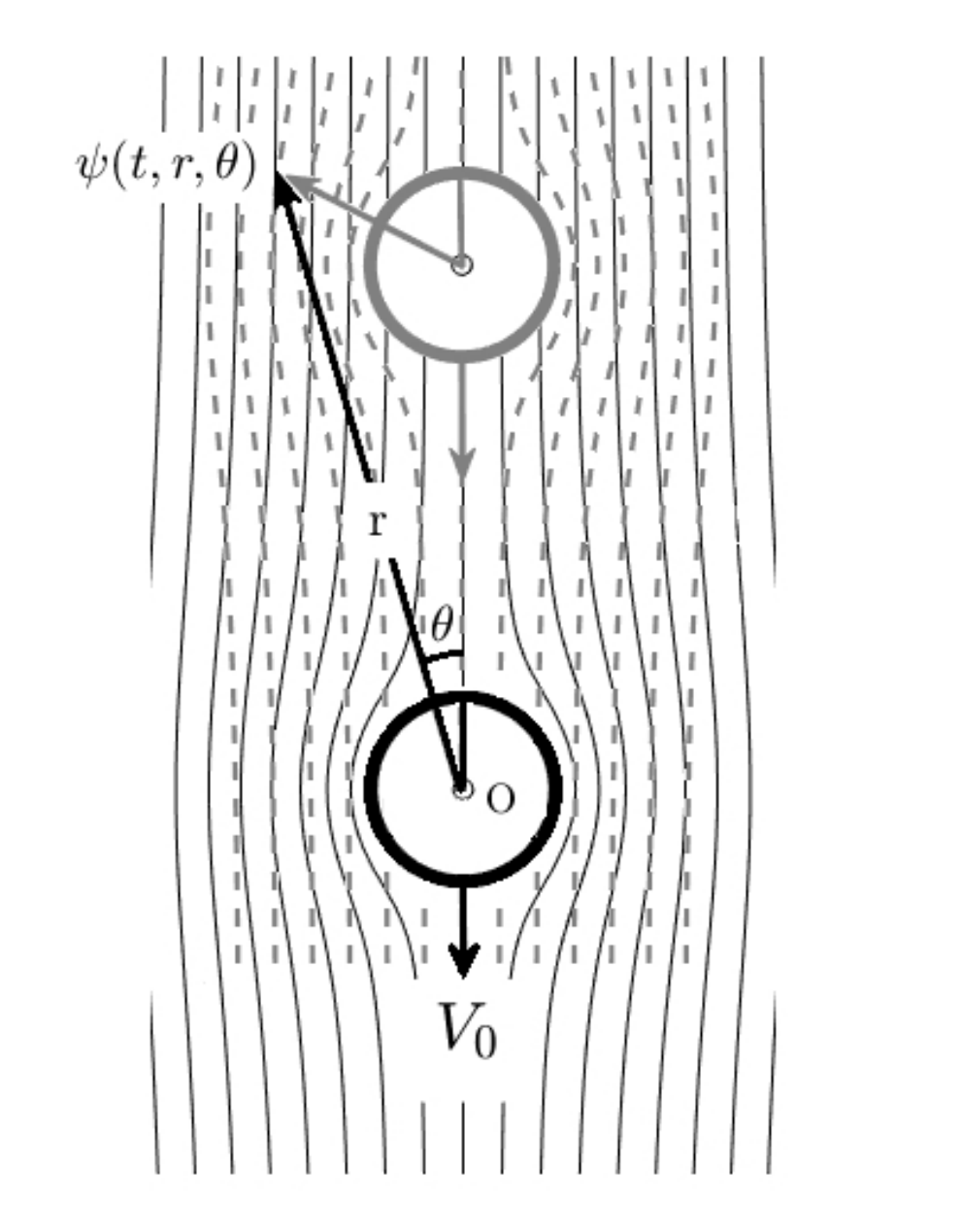}
    \caption{The laminar stream function around a sphere falling with constant velocity $V_0$ at an earlier (dashed top) and a later time $t$ (solid bottom). The stream function changes in spatial co-ordinates $r$ and $\theta$ with respect to a stationary observer so that it is a function of both the sphere velocity and time.}
    \label{fig:stream_function_Stokes}
\end{figure}

\cite{stokes1850effect} considered a sphere of radius $a$ falling at constant velocity $V_0$ under gravity along a straight axis $z$, considering the center of the sphere as the origin so that the motion of the fluid is symmetrical with respect to the axis of fall. Relative to the center of sphere, in a spherical coordinate system $(r,\theta,\phi)$ where $r$ is the radius, $\theta$ is the zenith angle, and $\phi$ is the azimuthal angle (Fig. \ref{fig:stream_function_Stokes}), the $u_r$ and $u_{\theta}$ components of velocities along and perpendicular to the direction of $r$ are

\begin{align}
    v_r(t,r,\theta) &= \frac{1}{r^2 \mathrm{sin}\theta} \pd{\psi(t,r,\theta)}{\theta}
    \label{eq:stream-function1} \\
    v_{\theta}(t,r,\theta) &= -\frac{1}{r \mathrm{sin}\theta} \pd{\psi(t,r,\theta)}{r}
    \label{eq:stream-function2}
\end{align}

For an incompressible fluid in an azimuthally symmetric 2D spherical coordinate system, the Navier--Stokes equations that determine the surrounding fluid flow around the moving sphere are

\begin{align}
    \pd{v_r}{t} + v_r \pd{v_r}{r} + \frac{v_{\theta}}{r} \pd{v_r}{\theta} -\frac{v_{\theta}^2}{r} &= -\frac{1}{\rho_f} \pd{p}{r} +\nu \Big(\nabla^2 v_r -\frac{2 v_r}{r^2} -\frac{2}{r^2} \pd{v_{\theta}}{\theta} -\frac{2 v_{\theta} cot\theta}{r^2} \Big)
    \label{eq:Navier--Stokes1} \\
    \pd{v_{\theta}}{t} + v_r \pd{v_{\theta}}{r} + \frac{v_{\theta}}{r} \pd{v_{\theta}}{\theta} +\frac{v_r v_{\theta}}{r} &= -\frac{1}{r \rho_f} \pd{p}{\theta} +\nu \Big(\nabla^2 v_{\theta} +\frac{2}{r^2} \pd{v_r}{\theta} -\frac{ v_{\theta}}{r^2 \mathrm{sin}^2\theta} \Big) \label{eq:Navier--Stokes2}
\end{align}

Assuming the no-slip condition at the surface of sphere, at a constant fall velocity $V_0$ the boundary conditions are

\begin{equation}
    v_r \bigg |_{r=a} = V_0 \ \mathrm{cos}\theta, \quad v_\theta \bigg |_{r=a} = -V_0 \ \mathrm{sin}\theta
\end{equation}

Defining for brevity an operator

\begin{equation}\label{eq:operator}
    \mathrm{D} = \pd[2]{}{r^2} + \frac{1}{r^2} \pd[2]{}{\theta^2} - \frac{\mathrm{cos}\theta}{r^2 \mathrm{sin}\theta} \pd{}{\theta} 
\end{equation}

then using \eqref{eq:stream-function1}-\eqref{eq:stream-function2}, the Navier--Stokes equations \eqref{eq:Navier--Stokes1}-\eqref{eq:Navier--Stokes2} can be rewritten in terms of the stream function as follows

\begin{align}
    -\frac{1}{\rho_f} \pd{p}{r} &= \frac{1}{r^2 \mathrm{sin}\theta} \pd{}{\theta} \Big(\pd{\psi}{t} -\nu \mathrm{D} \psi \Big)
    \label{eq:Navier--Stokes-stream-function1} \\
    \frac{1}{\rho_f} \pd{p}{\theta} &= \frac{1}{\mathrm{sin}\theta} \pd{}{r} \Big(\pd{\psi}{t} -\nu \mathrm{D} \psi \Big)
    \label{eq:Navier--Stokes-stream-function2}
\end{align}

Taking the derivative of \eqref{eq:Navier--Stokes-stream-function1} with respect to $\theta$ and the derivative of \eqref{eq:Navier--Stokes-stream-function2} with respect to $r$ and eliminating the pressure term, the equation for $\psi(t,r,\theta)$ becomes

\begin{equation} \label{eq:stream-function-equation}
    \underbrace{ \mathrm{D} \Big(\nu \mathrm{D}- \pd{}{t} \Big) \psi}_{linear} + \ \mathrm{sin}\theta \underbrace{ \bigg( \pd{\psi}{r} \pd{}{\theta} - \pd{\psi}{\theta} \pd{}{r} \bigg) \frac{\mathrm{D \psi} }{r^2 \mathrm{sin}^2\theta} }_{non-linear} = 0
\end{equation}

The solution to \eqref{eq:stream-function-equation} is the stream function for a viscous and incompressible fluid surrounding a moving sphere. Note the distinction between the linear term that produces a laminar flow around a slow-moving sphere and the non-linear term that arises from retaining the velocity products and squares in \eqref{eq:Navier--Stokes1}-\eqref{eq:Navier--Stokes2}.

In 1850, Stokes solved the linear term at steady-state, namely $\mathrm{D} \big(\mathrm{D} \psi(r,\theta) \big) = 0$ by switching reference frames and treating the fluid as moving with velocity $V_0$ relative to a stationary sphere. Therefore, by placing the origin at the center of the quiescent sphere, and supposing a solution in the form of $\psi(r,\theta) = \mathrm{sin}^2(\theta) f(r)$, \cite{stokes1850effect} determined the motion of a fluid for a sphere that moves slowly at a constant velocity in a fluid at rest

\begin{equation} \label{eq:Basset-infinity}
    \psi(r,\theta) = \frac{1}{4} V_0 a^2 \mathrm{sin}^2\theta \Big(\frac{3r}{a} - \frac{a}{r} \Big)
\end{equation}

Stokes obtained the familiar expression $F_D = 6 \pi \mu a V_0$ for the drag force of the fluid on the sphere assuming the no-slip condition at the sphere's surface. The terminal velocity of a falling sphere is then obtained from balance with the gravitational force (Appendix \ref{appA}).

\section{Overview of Basset's solution} \label{sec:Basset-solution}
Basset argued that Stokes' formula for the terminal velocity yields values larger than those obtained by experiment. Based on his prior theoretical studies \citep{basset1888treatise}, \cite{basset1910descent} attributed the discrepancy to the neglect of the $\pd{\psi}{t}$ term in \eqref{eq:stream-function-equation} for steady motion, suggesting that it should be replaced by $V_0 \pd{\psi}{z}$, and again maintaining the origin at the  center of the moving sphere (Fig. \ref{fig:stream_function_Stokes}). Stokes' assumption that sphere starts the motion with a constant velocity $V_0$ also implies a discontinuity at the sphere surface. Suppose that a sphere that is set in motion with a constant velocity of $V_0$. The no-slip condition requires that the fluid velocity instantly change from $\frac{1}{2} V_0 \ \mathrm{sin}\theta$ to $-V_0 \ \mathrm{sin}\theta$ (Appendix \ref{appB}). This discontinuity is unphysical. If instead the sphere is moving with a variable velocity $V(t)$ starting from rest then the revised linear equation to be solved is

\begin{equation} \label{eq:Eq.I}
    {\mathrm{D} \Big(\nu \mathrm{D} - \pd{}{t} \Big) \psi(t,r,\theta)}=0
\end{equation}

The solution was found first by \cite{boussinesq1885resistance,boussinesq1885applications} and apparently independently three years later by \cite{basset1888treatise}. Any more generalized analytical solution to \eqref{eq:stream-function-equation} has yet to be determined. 

Much has been written about the Basset drag force in the literature but less about how it was originally derived. Here, we revisit Basset's solution for two reasons. His work on the problem of variable slow motion of a sphere in a viscous fluid was last published in 1888, and the innovative analytical methods he used to solve partial differential equations are not well known. Second, we extend his mathematical approach to present a revised form of the Maxey--Riley equation suitable for application to a wider range of Reynolds numbers than the Stokes regime.

The solution to \eqref{eq:Eq.I} is outlined in more detail in Appendix \ref{appA}. Briefly, Basset's approach to solving \eqref{eq:Eq.I} for $\psi(t,r,\theta)$ was motivated by the absence of an analytical solution to the linear form of the Navier--Stokes equation for an accelerating particle. He began by first assuming that the sphere moves with constant velocity $V_0$. In this case, the particular solution for the stream function around a sphere with a moving origin \eqref{eq:Eq.I} is

\begin{align} \label{eq:Basset-solution0}
    \psi(t,r,\theta) =& \frac{1}{2} V_0 a^2 \ \mathrm{sin}^2\theta \Big\{\frac{3 \nu t}{r a} + \frac{6 \sqrt{\nu t/\pi}}{r} + \frac{a}{r} \Big\} \\
    &-\frac{3}{\sqrt{\pi}} V_0 a^2 \ \mathrm{sin}^2\theta \int_{\frac{r-a}{2 \sqrt{\nu t}}}^{\infty} \Big\{\frac{2 \xi^2 \nu t}{ra} + \frac{2 \xi \sqrt{\nu t}}{r} + \frac{1}{2}(\frac{a}{r} - \frac{r}{a})\Big\} e^{-\xi^2} d\xi
    \nonumber
\end{align}

The stream function around the sphere obtained by the Basset \eqref{eq:Basset-solution0} is laminar and its form is identical to that obtained by Stokes \eqref{eq:Basset-infinity}, as shown in Fig. \ref{fig:stream_function_Stokes}. The  difference is that the stream function is non-steady due to acceleration of the fluid around the sphere. Basset's unsteady stream function reduces to the Stokes steady stream function at the particle surface $r=a$, and in the limit $t\rightarrow\infty$ where the integral term of Basset's solution with ${r-a}/{2 \sqrt{\nu t}}$ approaches zero. At a distance radially far from the particle surface, or for shorter times where the fluid has not yet reached a steady motion, the value of stream function calculated by Basset's solution is greater than that found by Stokes.

By substituting \eqref{eq:Basset-solution0} into \eqref{eq:Navier--Stokes-stream-function1}, the solution for the fluid pressure field is 

\begin{equation} \label{eq:Basset-pressure}
    p(t,r,\theta) = \frac{3 V_0 a \mu \ \mathrm{cos}\theta}{2 r^2} \big(1 + \frac{a}{\sqrt{\pi \nu t}} \big)
\end{equation}
and the fluid velocities $v_r$ and $v_{\theta}$ are obtained by substituting \eqref{eq:Basset-solution0} into \eqref{eq:stream-function1} and \eqref{eq:stream-function2}, respectively. 

The drag force owes to the upstream pressure gradient across a falling particle and the shear stress in the particle boundary layer. At the sphere surface where $r=a$, the drag force is

\begin{align} \label{eq:Basset-drag} \nonumber
    F_D = 2 \pi a^2 \int_{0}^{\pi} \bigg\{ \Big(p - 2 \mu \pd{v_r}{r} \Big) \mathrm{cos}\theta + \mu \Big(\pd{v_\theta}{r} + \frac{1}{r} \pd{v_r}{\theta} -\frac{v_\theta}{r} \Big) \mathrm{sin}\theta \bigg\} \mathrm{sin}\theta \ d\theta \\
    = 6 \pi \mu a V_0 \big(1+ \frac{a}{\sqrt{\pi \nu t}} \big)
\end{align}

Neglecting velocity squared terms, there is a correction term to the Stokes drag. For physical insight, suppose that there is a relaxation time to the terminal velocity $\tau_p= V_0/g$ that in the Stokes flow regime is equal to $\tau_p = {m_p}/{6 \pi \mu a}$, simplifying to $\tau_p = \frac{ \rho_p d^2_p}{18 \mu}$. The fractional addition to the Stokes drag in \eqref{eq:Basset-drag} varies temporally as ${a}/{\sqrt{\pi \nu t}}$, which is proportional to $\sqrt{{\tau_p}/{t}}$. Then, $\tau_p$ is the time $t_{max}$ at which the unsteady drag is a maximum. Fluid accelerations around the particle surface exert a force on the particle that is proportional to the particle cross-section. The perturbation diffuses away from the particle as $1/\sqrt{t}$. For the case of turbulent flows, it has been suggested that the appropriate timescale to which particle relaxation time could be compared is the Kolmogorov timescale $\tau_\eta$ where $\eta$ is the Kolmogorov length scale, in which case the fractional enhancement of unsteady drag to Stokes drag is $\sim a/\eta$ \citep{daitche2015role}.

Effectively then, there is an extra drag force at constant $V_0$ that prolongs the time it takes the particle to approach its terminal velocity. For the more physical case that $V$ is not a constant, Basset's approach was to substitute in \eqref{eq:Basset-drag} the time variable $t$ with a historical time $\tau$, and $V_0$ with a time-varying velocity of form $\frac{dV(t-\tau)}{dt} d\tau$, integrating the result from $0$ to $t$. To see the justification for this substitution, consider that the transformation $\zeta = t-\tau$ leads to

\begin{align} \label{eq:velocity_variation}
    \int_{0}^{t} \frac{dV(t-\tau)}{dt} d\tau &= - \int_{t}^{0} \frac{dV(\zeta)}{d\zeta} d\zeta \\
    \nonumber
    &= V(t) - V(0)
\end{align}

If the sphere starts from rest, then  $V(0) = 0$, and $V(\zeta)$ is finite between its limits, any integration of a time varying velocity in \eqref{eq:velocity_variation} will yield the current sphere velocity at time $t$. \cite{basset1888treatise} proposed that if $V(t)$ is a solution to a partial differential equation, then the integral of $\frac{d V(t-\tau)}{dt} d\tau$ must also be a solution. The total drag force in \eqref{eq:Basset-drag} then becomes

\begin{equation} \label{eq:Basset-drag-timechange}
    F_D = 6 \pi \mu a \Big( V(t) + a \int_{0}^{t} \frac{1}{\sqrt{\pi \nu \tau}} \frac{dV(t-\tau)}{dt} \ d\tau \Big)
\end{equation}

Drag is not only a function of the current velocity but also of the particle acceleration due to prior interactions between the particle and the fluid.  \cite{basset1910ondescent} later adopted a method developed by \cite{picciati1907sul} that simplifies the procedure of first find a solution for constant velocity and then for changing velocity. Picciati's method reduces the problem to the determination of a function that satisfies Fourier's heat equation, and yields a solution equivalent to \eqref{eq:Basset-drag-timechange}. The equation of motion for a sphere of mass $m_p$ moving slowly with a time-varying velocity becomes

\begin{align} \label{eq:Basset-motion-equation}
    m_p \frac{d \textit{\textbf{V}}(t)}{dt} = (m_p - m_f) \textit{\textbf{g}} - 6 \pi \mu a \Big( \textit{\textbf{V}}(t) + a \int_{0}^{t} \frac{1}{\sqrt{\pi \nu (t-\tau)}} \frac{d\textit{\textbf{V}}(\tau)}{d\tau} \ d\tau \Big) - \frac{1}{2} m_f \frac{d\textit{\textbf{V}}(t)}{dt}
\end{align}

Equation \eqref{eq:Basset-motion-equation} does not consider the squares and products of flow velocities in the Navier--Stokes equation \eqref{eq:Navier--Stokes1}-\eqref{eq:Navier--Stokes2} and so it remains valid only for Stokes flow. It is this equation of motion that \citep{Tchen1947mean} employed to account for the effects of temporal variability in the fluid flow and that with subsequent revisions led to the Maxey--Riley equation of motion \eqref{eq:Maxey--Riley}.

\section{Unsteady drag at high Reynolds numbers}
To determine the hydrodynamic fluid forces at higher Reynolds numbers, what is required is a particular solution to the full Navier--Stokes equations \eqref{eq:Navier--Stokes1}-\eqref{eq:Navier--Stokes2}. This is not yet possible due to the mathematical difficulties introduced when higher order velocity terms are retained. Consequently, these terms have traditionally been either ignored or parameterized based on empirical studies. In the latter case, the steady drag on a falling sphere with velocity $V_0$ in a stationary and incompressible viscous fluid can then be expressed using Rayleigh’s formula $F_d = \frac{1}{2} \rho_f A_p C_D(\Rey) V^2_0$, in which case the drag force becomes

\begin{equation} \label{eq:Basset-drag0}
    F_D = \frac{1}{2} \rho_f A_p C_D(\Rey){V_0}^2 \Big(1 + \frac{a}{\sqrt{\pi \nu t}} \Big)
\end{equation}

For example, in the Stokes flow regime, a formulation for the drag coefficient $C_D(\Rey) = {24}/{\Rey}$ converts the Rayleigh formula to the familiar expression $F_d = 6 \pi \mu a V_0$ expressed in \eqref{eq:Basset-drag}. For higher Reynolds numbers, empirical estimates of the drag coefficient can be used. 

But if a more generalized drag force is to be implemented within the context of an equation such as the Maxey--Riley equation, appropriate adjustments must be made to the equation itself. We now proceed to derive an expression for the unsteady drag at high Reynolds numbers in a manner analogous to that described in Section \ref{sec:Basset-solution} for low Reynolds numbers. Following Basset's approach leading to \eqref{eq:Basset-drag-timechange} by way of \eqref{eq:velocity_variation}, a more general equation of motion is then

\begin{multline} \label{eq:new-motion-equation}
    m_p \frac{d V(t)}{dt} = (m_p - m_f) g -\frac{1}{2} \rho_f A_p \bigg( C_D(\Rey) V^2(t) + a \int_{0}^{t} \frac{C_D(\tau) |V(\tau)|}{\sqrt{\pi \nu(t-\tau)}} \frac{dV(\tau)}{d\tau} d\tau \bigg) \\
    -\frac{1}{2} m_f \frac{dV(t)}{dt}
\end{multline}

where the integral term expresses a more generalized unsteady drag. A possible limitation of this expression is that $C_D$ is derived empirically for a particle moving at constant velocity, and does not account for any time variation in the drag due to acceleration. Experimental studies suggest drag coefficients that can be significantly higher \citep{hughes1952er,selberg1968drag,igra1993shock}. 

What is important to note however is that within the integrand in \eqref{eq:new-motion-equation}, the particle acceleration is multiplied by the magnitude of the particle velocity, whereas with the Basset equation \eqref{eq:Basset-motion-equation}, it is multiplied by a constant. Therefore, for a particle falling at high velocity with a large Reynolds number, unsteady drag is not necessarily negligible as has sometimes been assumed. Ignoring any alterations to the drag force due to forced velocity curvature around the sphere (the Faxén correction $\frac{1}{6} a^2 \nabla^2 u_f$), and viscous shear stress, we propose a more general version of the Maxey--Riley equation of particle motion

\begin{align}
    m_p\frac{d\textit{\textbf{v}}_p}{dt} =& (m_p-\rho_f V_p)\textit{\textbf{g}} + \rho_f V_p\frac{D\textit{\textbf{v}}_f}{Dt}
    - \frac{1}{2} \rho_f A_p \Big\{C_D(\Rey) |\textit{\textbf{v}}_p-\textit{\textbf{v}}_f| (\textit{\textbf{v}}_p-\textit{\textbf{v}}_f)
    \nonumber \\
    & + a \int_{0}^{t} \frac{C_D(\tau) |\textit{\textbf{v}}_p(\tau)-\textit{\textbf{v}}_f(\tau)|}{\sqrt{\pi \nu (t-\tau)}} \frac{d\big(\textit{\textbf{v}}_p(\tau)-\textit{\textbf{v}}_f(\tau)\big)}{d\tau} \ d\tau \Big\} - k\rho_f V_p\frac{d}{dt} (\textit{\textbf{v}}_p-\textit{\textbf{v}}_f)
    \label{eq:modified-Maxey--Riley}
\end{align}

where $\Rey = |v_p(t) - v_f(t)| d_p/\nu$ is the particle's relative Reynolds number.

\section{Numerical analysis}
The equation of motion \eqref{eq:new-motion-equation} is now solved numerically. The particle velocity is initialized at some value close to zero, the particle's Reynolds number $\Rey={|V(t)|d_p}/{\nu}$ is specified, and the drag coefficient is calculated from an empirically derived relationship between the drag coefficient and the Reynolds number of a rigid sphere \citep{whiteviscous}

\begin{equation}\label{eq:drag_coefficient}
    C_D(Re)=\begin{cases}
    \frac{1}{4} + \frac{24}{Re} +\frac{6}{1+\sqrt Re} & \ \ \ \text{if $\Rey \leq 3,000$}\\
    0.3659, & \ \ \ \ \ \ \text{otherwise}
    \end{cases}
\end{equation}

The history term in \eqref{eq:new-motion-equation} can be estimated except where $\tau$ approaches $t$, at which point the integrand becomes infinite and must be treated separately. Using the definition of an integral, the history term evolves from the previous time step $(t-\Delta{t})$ through

\begin{align} \label{eq:Basset_integral}
  \int_{0}^{t} \frac{V(\tau)}{\sqrt{t-\tau}} \frac{dV(\tau)}{d\tau} \ d\tau &= \lim_{n\rightarrow\infty}\sum_{k=1}^{n-1} \bigg(\frac{V(t-k\Delta{t})}{\sqrt{k}} \ \frac{dV(t-k\Delta{t})}{dt} \bigg) \cdot \Delta{t}^{1/2}
\end{align}

where $t$ is the time of motion and $\Delta{t}$ is the time interval employed in the simulation. The right-hand side of \eqref{eq:Basset_integral} is amenable to standard numerical techniques.

\begin{figure}
    \centering
    \includegraphics[width=14cm, height=8cm]{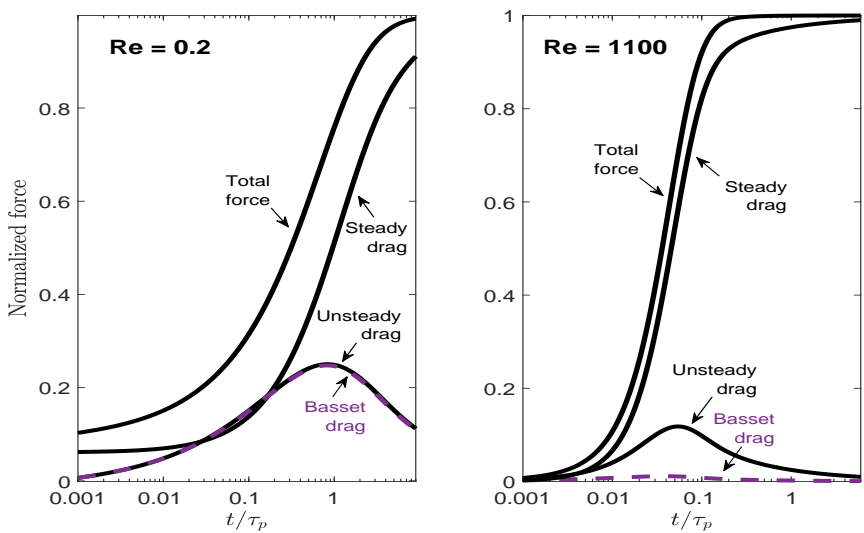}
    \caption{Normalized steady and unsteady drag forces acting on a particle falling into stationary air over a normalized log-time of motion, according to \eqref{eq:new-motion-equation}. The forces are normalized by gravity and the time of motion by the particle relaxation time in Stokes flow $\tau_p$. The normalized total force is also shown. Left) particle with a density ratio of $s=15$ and a low Reynolds number of $\Rey = 0.2$ Right) particle with a density ratio of $s=830$ and a Reynolds number of $\Rey = 1100$. The dashed purple line shows for comparison the unsteady Basset drag calculated using the Maxey--Riley equation of motion \eqref{eq:Maxey--Riley}.}
    \label{fig:norm_drag_force}
\end{figure}

Equation \eqref{eq:new-motion-equation} was solved numerically for the approach of a particle initially at rest to its terminal velocity considering both a low and high Reynolds number. Fig. \ref{fig:norm_drag_force} shows a comparison of steady and unsteady drag forces normalized by the gravity force as a function of time normalized by the particle Stokes time $\tau_p$. For a particle with a Reynolds number of $\Rey = 0.2$ the generalized equation for unsteady drag, the integral term in \eqref{eq:new-motion-equation}, is equivalent to the Basset history term and is a maximum $25\%$ of the total force when the Stokes time $t/\tau_p \simeq 1$. Its contribution to the particle acceleration is negligible as the drag turns steady and the particle approaches its terminal velocity. 

\begin{figure}
    \centering
    \includegraphics[width=9cm, height=8cm]{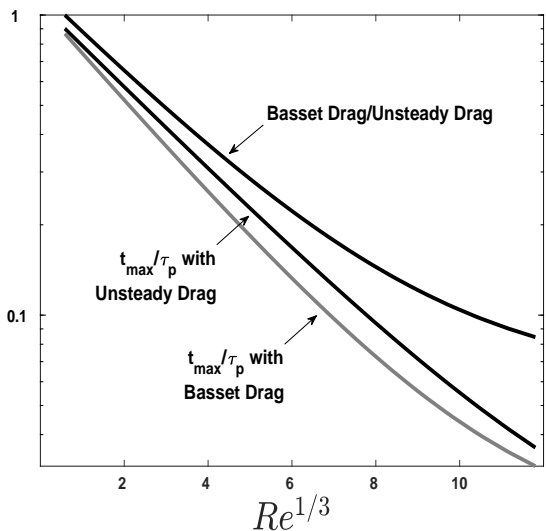}
    \caption{Time relative to the Stokes time $\tau_p$ at which the unsteady drag is a maximum, and the corresponding ratio of the Basset drag to the unsteady drag, as a function of the cube root of the Reynolds number.}
    \label{fig:drag_different_Re}
\end{figure}

For a higher Reynolds number of $\Rey = 1100$, the unsteady drag accounts for a maximum $\sim15\% $ of the total force at a time much shorter than particle relaxation time $\tau_p$ while the Basset history drag plays a negligible role. Fig. \ref{fig:drag_different_Re} shows that the time at which the unsteady drag reaches its maximum value decreases logarithmically as the cube root of the particle Reynolds number $\Rey^{1/3}$, or that $\ln(t_{max}/\tau_p) \propto -\Rey^{1/3}$. Also shown is the ratio of the Basset drag to the generalized unsteady drag, which also decreases as $\Rey^{1/3}$, indicating a diminished relative importance of the Basset drag at higher Reynolds numbers. So while the revised unsteady drag dominates the Basset history drag, and consequently increases the drag and reduces the particle terminal velocity, relative to the Stokes time, the period over which the drag affects the particle motion is correspondingly short.

\section{Discussion}
There remain some important limitations to \eqref{eq:new-motion-equation}. First, there is an implicit assumption that the particle starts from rest. While nonetheless assuming Stokes flow, \cite{basset1888treatise} developed a rather more complicated equation of motion for a particle initially projected vertically with velocity $V_i$ (for derivation, see Appendix \ref{appB})

\begin{align} \label{eq:Basset-motion-equation-inivelocity}
    \frac{d \textit{\textbf{V}}(t)}{dt} = - \lambda \textit{\textbf{V}}_i e^{-\lambda t} + f \textit{\textbf{g}} \ e^{-\lambda t} - \lambda a \ \frac{d}{dt} \int_{0}^{t} \int_{0}^{\upsilon} \frac{e^{-\lambda (t-\upsilon)}}{\sqrt{\pi \nu (\upsilon-\tau)}} \frac{d\textit{\textbf{V}}(\tau)}{d\tau} \ d\tau \ d\upsilon
\end{align}

The coefficient $f = \frac{m_p - m_f}{m_p + \frac{1}{2} m_f}$ simplifies to $f = \frac{\rho_p - \rho_f}{\rho_p + \frac{1}{2} \rho_f}$, and $\lambda = \frac{6\pi \mu a}{m_p + \frac{1}{2} m_f}$ to $\lambda = \frac{9 \rho_f \nu}{2 a^2 (\rho_p + \frac{1}{2}\rho_f)}$. Basset was unable to integrate this complicated integro-differential equation, but for the limited case of $\lambda \ll 1$, as applies to a sphere moving in a fluid whose kinematic viscosity is small, he used a method of successive approximation to obtain the acceleration and velocity to the third power in $\lambda$. 

Later, \cite{boggio1907integrazione} successfully reduced the complexity of the problem to a solvable second order differential equation (see Appendix \ref{appC}). The solution employs error functions of form $\mathrm{erf(\sqrt{\alpha t})}$ and $\mathrm{erf(\sqrt{\beta t})}$ where $\alpha,\beta = \frac{\lambda}{2}\{(q-2) \pm \sqrt{q(q-4)}\}$ and $q={\lambda a^2}/{\nu}$. Substituting this expression for $\lambda$ yields $q=\frac{9 \rho_f}{2\rho_p + \rho_f}$. For a particle denser than the fluid then $q<4$, and $\alpha$ and $\beta$ are complex numbers. For this case, 

\begin{align} \label{eq:Basset-analytical-solution-inivelocity}
    V(t) = \frac{f g}{\lambda} &+ \big(V_i - \frac{f g}{\lambda} \big) e^{\gamma t} \big\{\mathrm{cos}(\delta t) - \frac{\gamma + \lambda}{\delta} \ \mathrm{sin}(\delta t)\big\} \\
    &-\frac{h \ e^{\gamma t}}{\delta} \bigg\{\mathrm{cos}(\delta t) \int_{0}^{t} \frac{e^{-\gamma t} \mathrm{sin}(\delta t)}{\sqrt{t}} dt- \mathrm{sin}(\delta t) \int_{0}^{t} \frac{e^{-\gamma t} \mathrm{cos}(\delta t)}{\sqrt{t}} dt \bigg\}
    \nonumber
\end{align}

where $\gamma = \frac{\lambda}{2}(q-2) = - \frac{\lambda}{2} \big(\frac{4\rho_p-7\rho_f}{2\rho_p+\rho_f}\big)$, $\delta = \frac{\lambda}{2} \sqrt{q(4-q)} = \frac{\lambda}{2} q^{\frac{1}{2}} \sqrt{\frac{8\rho_p-5\rho_f}{2\rho_p+\rho_f}}$, and $h = \frac{\lambda a}{\sqrt{\pi \nu}} (f g - \lambda V_i)$. This equation is not widely known but it significantly reduces the computational expense of finding a solution for $V(t)$ by eliminating the requirement of tracking the history of the particle's motion. 

A second, more troubling limitation of \eqref{eq:Basset-motion-equation-inivelocity}-\eqref{eq:Basset-analytical-solution-inivelocity}, and hence also of \eqref{eq:Basset-motion-equation} and \eqref{eq:new-motion-equation}, is that for a particle starting at $t = 0$ with a finite vertical velocity, the effect of the initial velocity (or any disturbance to the flow field surrounding the sphere) on the eventual particle displacement does not decay to zero at infinite time. The end result is that the terminal velocity differs from that expected from the Stokes solution. While the effect is small, it nonetheless implies the unphysical property of infinite memory in a dissipative viscous fluid \citep{reeks1984dispersive}.

To resolve this issue, \cite{sano1981unsteady} applied a matching procedure initially developed by \cite{bentwich1978unsteady} to unsteady low Reynolds number flow past a sphere to find that the drag decays faster than $t^{-1/2}$ when $t \gg \tau_p$. Thus, the temporal dependence of the Basset drag is only appropriate at times less than $\tau_p$ when inertial forces are low compared to viscous forces. A similar conclusion was reached by \cite{mei1991unsteady}. \cite{mei1992flow} applied a successive orders of approximation method to solve the Navier--Stokes equation to $\mathrm{O}(\mathrm{Re})$ for the case of oscillating flow over a sphere, by considering small fluctuations in velocity when the Reynolds number is not negligibly small. \cite{mei1992flow} then proposed a modified expression for the unsteady drag that includes an integration kernel that decays as $t^{-2}$ for $t\gg \tau_p$, limited to finite Reynolds numbers $(\mathrm{Re} \leq 100)$ and small-amplitude fluctuations in the velocity of the free stream. \cite{mei1994flow} later investigated the applicability of the kernel for other types of unsteady flows.

\cite{mainardi1997fractional} went further to interpret the Basset force in terms of a fractional derivative of any order $\ell$ ranging in the interval $0 < \ell < 1$ as

\begin{equation} \label{eq:Basset-drag-fractionalDeriv}
    F_D = \frac{9}{2}m_f \Big( \frac{1}{\tau_0} + \frac{1}{\tau^{1-\ell}_0} \frac{d^{\ell}}{{dt}^{\ell}} \ \Big) V(t)
\end{equation}

where $\tau_0= a^2/\nu$ represents the characteristic time to reach steady-state in a viscous fluid. $\ell = 1/2$ yields the total Basset drag in \eqref{eq:Basset-drag-timechange}. This generalization, suggested by mathematical speculation, modifies the behaviour of the solution, changing its decay from $t^{-1/2}$ to $t^\ell$ for $t\gg\tau_p$. \cite{mainardi1997fractional} considered three cases of $\ell = 1/4$, $1/2$, and $3/4$ and compared the particle terminal velocity with a desired temporal adjustment behavior $e^{-t/\tau_p}$ expected from Stokes drag $(\ell = 0)$. The results yielded improved agreement with the Stokes solution but the topic is still considered unsolved, as ideally it requires a full solution to the Navier--Stokes equations, including non-linear inertia terms involving the products of velocities.

\section{Conclusions}
The Maxey--Riley equation was originally developed for the study of small, slow-moving spheres but is widely used for higher Reynolds numbers under the assumption that unsteady Basset drag is insignificant relative to the steady drag. Here we have presented a historical review of the derivation of the equation of motion that leads to the Maxey--Riley equation and argue that the Basset drag can be suitably applied only when Reynolds numbers are small. Following Basset's original approach, but considering drag proportional to the particle relative velocity squared, a revised analytical equation is developed for extension to higher Reynolds numbers. Simulations based on this equation show that the unsteady drag force contributes substantially to the total drag at timescales less than the Stokes time, even for high values of the Reynolds number.

\section*{Acknowledgments}
This work is supported by the U.S. Department of Energy (DOE) Atmospheric System Research program award number DE-SC0016282 and the National Science Foundation (NSF) Physical and Dynamic Meteorology program award number 1841870.

\appendix \section{Basset's solution} \label{appA}
\cite{basset1888treatise} assumed a sphere of radius $a$ moving slowly in a fluid with a uniform velocity $V_0$, with its center at the origin moving in a straight line, surrounded by a viscous liquid that is initially at rest. As described in Section \ref{sec:Basset-solution}, the stream function must satisfy the linearized Navier--Stokes equation

\begin{equation} \label{eq:Eq.I-Append}
    {\mathrm{D} \Big(\mathrm{D} - \frac{1}{\nu} \pd{}{t} \Big) \psi} = 0
\end{equation}

where in a spherical coordinate system $(r,\theta,\phi)$ 

\begin{equation}
    \mathrm{D} = \pd[2]{}{r^2} + \frac{1}{r^2} \pd[2]{}{\theta^2} - \frac{\mathrm{cos}\theta}{r^2 \mathrm{sin}\theta} \pd{}{\theta} 
\end{equation}

Since the operators $\mathrm{D}$ and $(\mathrm{D} - \frac{1}{\nu} \pd{}{t})$ are commutative, following \cite{stokes1850effect}, Basset's solution to \eqref{eq:Eq.I-Append} was \citep{basset1888treatise}:

\begin{equation}
    \psi(t,r,\theta) = \psi_1(t,r,\theta) + \psi_2(t,r,\theta)
\end{equation}

where $\psi_1(t,r,\theta)$ and $\psi_2(t,r,\theta)$ satisfy, respectively

\begin{align} \label{eq:Eq.I-simplified}
    \begin{cases}
    \mathrm{D}\psi_1=0 \\
    \big(\mathrm{D}-\frac{1}{\nu} \pd{}{t} \big)\psi_2=0
    \end{cases}
\end{align}

Assuming the no-slip condition at the surface of a rigid falling sphere, the boundary conditions at the surface of sphere satisfying \eqref{eq:stream-function1}-\eqref{eq:stream-function2} are

\begin{equation} \label{eq:Boundary-conditions-stream}
    \pd{\psi}{\theta} \bigg |_{r=a} = V_0 a^2 \mathrm{sin}\theta \ \mathrm{cos}\theta, \quad \pd{\psi}{r} \bigg |_{r=a} = V_0 a \ \mathrm{sin}^2\theta
\end{equation}

It is important to mention that this last boundary condition shows that $\theta$ must appear in $\psi(t,r,\theta)$ in the form of the factor $\mathrm{sin}^2\theta$. Basset used separation of variables $\psi(t,r,\theta)=\psi(t,r)\mathrm{sin}^2\theta$ satisfying \eqref{eq:Eq.I-simplified}, to obtain

\begin{align}
    \pd[2]{\psi_1}{r^2}-\frac{2 \psi_1}{r^2}&=0 \label{eq:II-seperation} \\
    \pd[2]{\psi_2}{r^2}-\frac{2 \psi_2}{r^2}&=\frac{1}{\nu} \pd{\psi_2}{t} \nonumber
\end{align}

The particular solution of $\psi_1(t,r)$ is $f(t)/r$. In an innovative approach, Basset assumed a solution of form

\begin{equation}
    \psi_1 = \frac{1}{2 r}\sqrt{\frac{\pi}{\nu t}} \int_{0}^{\infty} \kappa(\rho) e^{-\rho^2/4 \nu t} \ d\rho
\end{equation}

Provided the solution satisfies the boundary conditions, there is no restriction on how $\psi_1$ varies with time. For this purpose Basset chose a Gaussian distribution $\kappa(\rho) e^{-\rho^2/4 \nu t}$. Note that $\rho$ has units of length. It is zero by definition at the particle surface and increases to infinity far from the particle. $\kappa(\rho)$ is an arbitrary function to be established.
To find the solution for $\psi_2$, Basset used separation of variables $\psi_2(t,r) = T(t) R(r)$ so that

\begin{equation}
    \frac{1}{R(r)} \frac{d^2 R(r)}{dr^2} -\frac{2}{r^2} = \frac{1}{\nu T(t)} \frac{d^2 T(t)}{dt^2}
\end{equation}

The LHS is strictly a function of $r$ and RHS of $t$, so for any range of time and space integration, both are equal and therefore can be assigned to an arbitrary constant $-m^2$ where the value of $m$ can be specified from a particular set of boundary conditions. Note that the solution for the functionality in $R(r)$ is spherical Bessel functions of the first and second kind that can be written in terms of Rayleigh's formulas

\begin{equation}
    R(r) = A r \frac{d}{dr} \frac{\mathrm{cos}(m r)}{r} + B r \frac{d}{dr} \frac{\mathrm{sin}(m r)}{r}
\end{equation}

With respect to the time functionality

\begin{equation}
    T(t) = C e^{-m^2 \nu t}
\end{equation}

Basset expressed a particular solution to $\psi_2$ in \eqref{eq:II-seperation} through use of boundary conditions \eqref{eq:Boundary-conditions-stream} at the particle surface

\begin{equation} \label{eq:psi_2}
    \psi_2 = A r \frac{d}{dr} \frac{e^{-m^2 \nu t}}{r} \mathrm{cos}[m(r-a+\rho)]
\end{equation}

Integrating \eqref{eq:psi_2} with respect to $m$ between the limits $0$ and $\infty$ to consider all the possible values of $m$, exchanging $A$ with $A(\rho)$, and integrating the results with respect to $\rho$ between the same limits yields:

\begin{equation}
    \psi_2 = \frac{r}{2} \sqrt{\frac{\pi}{\nu t}} \frac{d}{dr} \int_{0}^{\infty} \frac{A(\rho)}{r} e^{-\frac{(r-a+\rho)^2}{4 \nu t}} d\rho
\end{equation}

First differentiating and then integrating by parts $\int{}^{}udv=uv-\int{}^{}vdu$ where $u = A(\rho)/r$, Basset obtained

\begin{align}
    \psi_2 = \frac{r}{2} \sqrt{\frac{\pi}{\nu t}} \int_{0}^{\infty} -\frac{A(\rho)}{r^2}e^{-\frac{(r-a+\rho)^2}{4 \nu t}} d\rho + \Big [\frac{A(\rho)}{r} e^{-\frac{(r-a+\rho)^2}{4 \nu t}} \Big ]_{0}^{\infty} \nonumber \\
    - \frac{r}{2} \sqrt{\frac{\pi}{\nu t}} \int_{0}^{\infty} \frac{1}{r} \frac{dA(\rho)}{d\rho} e^{-\frac{(r-a+\rho)^2}{4 \nu t}} d\rho
    \nonumber
\end{align}

As $A(\rho)$ is a function of arbitrary form, it is possible to define it in such a way that the term in brackets can be eliminated at both limits, namely $A(0) = 0$ and $A(\rho) e^{-\rho^2} \rightarrow 0$ when $\rho \rightarrow \infty$. Thus, the total solution for $\psi(t,r,\theta)$ becomes

\begin{align} \label{eq:eq_stream_function0}
    \psi(t,r,\theta) =& \frac{\mathrm{sin}^2 \theta}{2r} \sqrt{\frac{\pi}{\nu t}} \int_{0}^{\infty} \kappa(\rho) e^{-\rho^2/4 \nu t} d\rho \\
    &-\frac{\mathrm{sin}^2 \theta}{2} \sqrt{\frac{\pi}{\nu t}} \int_{0}^{\infty} \Big \{\frac{A(\rho)}{r}+ \frac{dA(\rho)}{d\rho} \Big \} e^{-\frac{(r-a+\rho)^2}{4 \nu t}} d\rho 
    \nonumber
\end{align}

At the particle surface $r=a$, the exponential term in the second integral can be expressed in the form of the first integral. Thereby, the boundary condition $\pd{\psi}{\theta} \bigg |_{r=a} = V_0 a^2 \mathrm{sin}\theta \ \mathrm{cos}\theta$ is satisfied in \eqref{eq:eq_stream_function0} if

\begin{equation}
    \kappa(\rho)-A(\rho)-a \frac{dA(\rho)}{d\rho} = V_0 a^3/\pi
\end{equation}

Also, the boundary condition $\pd{\psi}{r} \bigg |_{r=a} = V_0 a \ \mathrm{sin}^2\theta$ requires that

\begin{align} \label{eq:integral}
    -\frac{1}{2} \sqrt{\frac{\pi}{\nu t}} \int_{0}^{\infty} \kappa(\rho) e^{-\rho^2/4 \nu t} d\rho +\frac{1}{2} \sqrt{\frac{\pi}{\nu t}} \int_{0}^{\infty} A(\rho) e^{-\rho^2/4 \nu t} d\rho \nonumber \\
    -\frac{a}{2} \sqrt{\frac{\pi}{\nu t}} \int_{0}^{\infty} \Big \{A(\rho)+ a \frac{dA(\rho)}{d\rho} \Big \} \frac{d}{d\rho} e^{-\rho^2/4 \nu t} d\rho = V_0 a^3
    \nonumber
\end{align}

Integrating by parts the last term on the LHS, and assuming that $[A(\rho)+a A'(\rho)] e^{-\rho^2/4 \nu t}$ disappears where $\rho=0$ and $\rho \rightarrow \infty$, what is required is that $A(0)=A'(0)=0$ and that $A(\rho) e^{-\rho^2/4 \nu t}$ and $A'(\rho) e^{-\rho^2/4 \nu t}$ should each vanish when $\rho \rightarrow \infty$. Then, equation \eqref{eq:integral} becomes

\begin{align}
    \frac{1}{2} \sqrt{\frac{\pi}{\nu t}} \int_{0}^{\infty} \Big\{-\kappa(\rho) + A(\rho) + a \frac{dA(\rho)}{d\rho} + a^2 \frac{d^2A(\rho)}{d\rho^2}\Big \} e^{-\rho^2/4 \nu t} d\rho = V_0 a^3
\end{align}

which is satisfied if

\begin{equation}
    -\kappa(\rho)+A(\rho)+a \frac{dA(\rho)}{d\rho} + a^2 \frac{d^2A(\rho)}{d\rho^2} = 2 V_0 a^3/\pi
\end{equation}

Hence, we have $A(\rho)=3 V_0 a \rho^2/2 \pi+ C_1 a + C_2$, where the condition $A(0)=A'(0)=0$ requires that $C_1=C_2=0$ and $\kappa(\rho)=V_0 a \pi^{-1}(\frac{3}{2}\rho^2+3\rho a + a^2)$. Then, the particular solution for the stream function \eqref{eq:eq_stream_function0} around a sphere moving with a uniform velocity $V_0$ is:

\begin{align}
    \psi(t,r,\theta) =& \frac{V_0 a \ \mathrm{sin}^2 \theta}{2r \sqrt{\pi \nu t}} \int_{0}^{\infty} (\frac{3}{2}\rho^2+3\rho a + a^2) e^{-\rho^2/4 \nu t} d\rho \\
    &-\frac{3 V_0 a \ \mathrm{sin}^2 \theta}{2 \sqrt{\pi \nu t}} \int_{0}^{\infty} \Big \{\frac{\rho^2}{2r} + \rho \Big \} e^{-\frac{(r-a+\rho)^2}{4 \nu t}} d\rho 
    \nonumber
\end{align}

The first term is in the form of an exponential integral, and the second can be solved through the substitution $r-a-\rho = 2 \zeta \sqrt{\nu t}$, from which Basset finally obtained

\begin{align} \label{eq:Basset-solution}
\psi(t,r,\theta) =& \frac{1}{2} V_0 a^2 \ \mathrm{sin}^2\theta \Big\{\frac{3 \nu t}{r a} + \frac{6 \sqrt{\nu t/\pi}}{r} + \frac{a}{r} \Big\} \\
    &-\frac{3}{\sqrt{\pi}} V_0 a^2 \ \mathrm{sin}^2\theta \int_{\frac{r-a}{2 \sqrt{\nu t}}}^{\infty} \Big\{\frac{2 \xi^2 \nu t}{ra} + \frac{2 \xi \sqrt{\nu t}}{r} + \frac{1}{2}(\frac{a}{r} - \frac{r}{a})\Big\} e^{-\xi^2} d\xi
    \nonumber
\end{align}

This is Basset's solution to \eqref{eq:Eq.I-Append}. It satisfies the Navier--Stokes equation \eqref{eq:Navier--Stokes1}-\eqref{eq:Navier--Stokes2} provided the advection terms involving velocity products and squares are omitted. At $t=0$, the integral vanishes in \eqref{eq:Basset-solution}, and the initial value of $\psi$ becomes

\begin{equation} \label{eq:Basset-zero}
    \psi(r,\theta) = \frac{V_0 a^3 \mathrm{sin}^2\theta}{2 r}
\end{equation}

which is the known solution of $\psi$ for the case of a frictionless liquid. When $t$ is very large, one may substitute $t \rightarrow \infty$ in the lower limit of the integral in \eqref{eq:Basset-solution} leading to

\begin{equation} \label{eq:stokes-solution}
    \psi(r,\theta) = \frac{1}{4} V_0 a^2 \mathrm{sin}^2\theta \Big(\frac{3r}{a} - \frac{a}{r} \Big)
\end{equation}

which is Stokes' solution for the motion of slowly moving rigid sphere in a viscous liquid after sufficient time has elapsed for the motion to become steady. One thing to note is that Stokes' steady-state solution for the stream function does not contain any expression of viscosity implying that the solution only applies to highly viscous liquids like water.

\section{Basset's equation of motion} \label{appB}
\cite{zeleny1910terminal} tested Stokes' formula for the terminal velocity of small spherical spores descending in the air under gravity, and they found that the value of terminal velocity calculated by resistance expressed by Stokes's solution yields values much larger than those obtained by their experiment. As mentioned in Section \ref{sec:Basset-solution}, \cite{basset1910descent} proposed that with respect to a moving origin, the term $\pd{\psi}{t}$ should be replaced by $V_0 \pd{\psi}{z}$. Although, Basset was unable to obtain a complete solution for steady motion due to the difficulty of obtaining  appropriate boundary conditions, his more general solution is nonetheless quite different from that given by Stokes \eqref{eq:stokes-solution}.

Stokes' solution also ignores inertia in the disturbed fluid flow, which alters the boundary conditions very far from the sphere \citep{oseen1910uber}. For a sphere that starts from rest in a stationary fluid, the hypothesis of no-slip condition holds at the surface of the sphere because both sphere and fluid have zero initial velocities. However, for a sphere that is set in motion with a constant velocity of $V_0$, there exists an initial motion of the fluid at the surface of the sphere due to an impulse needed to begin the sphere's motion. From the theory of impulsive motion, the fluid can be assumed to be frictionless at the beginning of the motion. The tangential velocity of the fluid from the equation for a frictionless fluid \eqref{eq:Basset-zero} becomes

\begin{equation}
    \Theta(r,\theta) = \frac{V_0 a^2 \mathrm{sin}\theta}{2 r^2}
\end{equation}

Which at the surface of sphere is $\frac{1}{2} V_0 \ \mathrm{sin}\theta$. The velocity components of the fluid along and perpendicular to the radius vector that satisfies the boundary conditions at the surface of the sphere applying the no-slip condition \eqref{eq:Boundary-conditions-stream} are

\begin{align}
    R(r,\theta) &= \frac{1}{a^2 \ \mathrm{sin}\theta} \pd{\psi}{\theta} = V_0 \ \mathrm{cos}\theta \\
    \Theta(r,\theta) &= - \frac{1}{a \ \mathrm{sin}\theta} \pd{\psi}{r} = - V_0 \ \mathrm{sin}\theta \nonumber
\end{align}

Therefore, the consequence of the no-slip condition at the surface of sphere is that the initial velocity of fluid suddenly changes from $\frac{1}{2} V_0 \ \mathrm{sin}\theta$ to $-V_0 \ \mathrm{sin}\theta$. This discontinuity of velocity has no physical interpretation. Although the initial motion of the fluid in the neighborhood of the sphere is highly turbulent and that it gradually subsides through the action of viscosity, but the consequence of no-slip condition is that the tangential velocity of the fluid is discontinuous at the surface of sphere. Basset proposed a simple procedure to suppose that the sphere is moving with a variable velocity $V(t)$ starting from rest.

As outlined in Section \ref{sec:Basset-solution}, \cite{basset1888treatise} developed the equation of motion for a sphere of mass $m_p$ that starts the motion from rest and then moves slowly with a time-varying velocity 

\begin{align} \label{eq:Basset-motion-equation1}
    \big(m_p + \frac{1}{2} m_f \big) \frac{d \textit{\textbf{V}}(t)}{dt} = (m_p - m_f) \textit{\textbf{g}} - 6 \pi \mu a \Big( \textit{\textbf{V}}(t) + a \int_{0}^{t} \frac{1}{\sqrt{\pi \nu (t-\tau)}} \frac{d\textit{\textbf{V}}(\tau)}{d\tau} \ d\tau \Big)
\end{align}

which can be simplified to

\begin{align} \label{eq:Basset-motion-equation2}
    \frac{d \textit{\textbf{V}}(t)}{dt} = f \textit{\textbf{g}} - \lambda \textit{\textbf{V}}(t) - \frac{\lambda a}{\sqrt{\pi \nu}} \int_{0}^{t} \frac{1}{\sqrt{t-\tau}} \frac{d\textit{\textbf{V}}(\tau)}{d\tau} \ d\tau
\end{align}

where coefficient $\lambda = \frac{6\pi \mu a}{m_p + \frac{1}{2} m_f}$ which thereby simplifies to $\lambda = \frac{9 \rho_f \nu}{2 a^2 (\rho_p + \frac{1}{2} \rho_f)}$, and $f = \frac{m_p - m_f}{m_p + \frac{1}{2} m_f}$. 

For a sphere set in motion with an initial velocity $V_0$, Basset divided the time $t$ into two intervals $\epsilon$ and $t - \epsilon$, where $\epsilon$ is a vanishing infinitesimal. In the first interval, the sphere starts to fall from rest due to gravity and a momentary external force that is large and constant. The value of the external force $(m_p + \frac{1}{2} m_f) X$ leads to a velocity $V_i$ at the end of the time interval so that $V_i = X\epsilon$. $X$ is the acceleration due to the momentary external force.

Replacing $f \textit{\textbf{g}}$ by $f \textit{\textbf{g}} + \textit{\textbf{X}}$, multiplying by $e^{\lambda t}$, and integrating between the limits $t$ and $0$ of \eqref{eq:Basset-motion-equation2}, Basset obtained

\begin{align} \label{eq:Basset-motion-equation-inivelocity1}
    \textit{\textbf{V}}(t) \ e^{\lambda t} = \int_{0}^{\epsilon} \textit{\textbf{X}} \ e^{\lambda \upsilon} d\upsilon + f \int_{0}^{t} \textit{\textbf{g}} \ e^{\lambda \upsilon} d\upsilon - \frac{\lambda a}{\sqrt{\pi \nu}} \int_{0}^{t} \int_{0}^{\upsilon} \frac{e^{\lambda \upsilon}}{\sqrt{\upsilon-\tau}} \frac{d\textit{\textbf{V}}(\tau)}{d\tau} \ d\tau \ d\upsilon
\end{align}

Note that integrating by parts $\int{}^{}udv=uv-\int{}^{}vdu$ where $u = e^{\lambda \upsilon}$ is applied to the LHS in \eqref{eq:Basset-motion-equation2}, which cancels the second term in the RHS. $d\textit{\textbf{V}}/dt$ consists of two components; a large one that is a function of \textit{\textbf{X}} equal to $\textit{\textbf{V}}_i/\epsilon$, and a second that is a function of $f \textit{\textbf{g}}$ and we continue to denote by $d\textit{\textbf{V}}/dt$. Hence equation \eqref{eq:Basset-motion-equation-inivelocity1} becomes

\begin{align} \label{eq:Basset-motion-equation-inivelocity2}
    \textit{\textbf{V}}(t) \ e^{\lambda t} = \frac{\textit{\textbf{X}}}{\lambda} (e^{\lambda \epsilon} - 1) + \frac{f \textit{\textbf{g}}}{\lambda} (e^{\lambda t} - 1) - \frac{\lambda a}{\sqrt{\pi \nu}} \int_{0}^{t} \int_{0}^{\upsilon} \frac{e^{\lambda \upsilon}}{\sqrt{\upsilon-\tau}} \bigg( \frac{d\textit{\textbf{V}}(\tau)}{d\tau} + \frac{\textit{\textbf{V}}_i}{\epsilon} \bigg) \ d\tau \ d\upsilon
\end{align}

The second part of integral becomes

\begin{align}
    \int_{0}^{\epsilon} \frac{2\textit{\textbf{V}}_i}{\epsilon} \upsilon^{1/2} e^{\lambda \upsilon} \ d\upsilon = 0, \ when \ \epsilon \rightarrow 0
\end{align}

Also, in the limit when $\epsilon$ vanishes, the first term in RHS of \eqref{eq:Basset-motion-equation-inivelocity2} becomes

\begin{align}
    \lim_{\epsilon \rightarrow 0} \frac{V_i}{\lambda \epsilon} (e^{\lambda \epsilon} - 1) = V_i
\end{align}

so the equation of motion becomes

\begin{align} \label{eq:Basset-motion-equation-inivelocity3}
    \textit{\textbf{V}}(t) = \textit{\textbf{V}}_i \ e^{-\lambda t} + \frac{f \textit{\textbf{g}}}{\lambda} (1-e^{-\lambda t}) - \frac{\lambda a}{\sqrt{\pi \nu}} \int_{0}^{t} \int_{0}^{\upsilon} \frac{e^{-\lambda (t-\upsilon)}}{\sqrt{\upsilon-\tau}} \frac{d\textit{\textbf{V}}(\tau)}{d\tau} \ d\tau \ d\upsilon
\end{align}

Note that the value of the acceleration is then

\begin{align} \label{eq:Basset-motion-equation-inivelocity4}
    \frac{d \textit{\textbf{V}}(t)}{dt} = -\lambda \textit{\textbf{V}}_i e^{-\lambda t} + f \textit{\textbf{g}} \ e^{-\lambda t} - \lambda a \ \frac{d}{dt} \int_{0}^{t} \int_{0}^{\upsilon} \frac{e^{-\lambda (t-\upsilon)}}{\sqrt{\pi \nu (\upsilon-\tau)}} \frac{d\textit{\textbf{V}}(\tau)}{d\tau} \ d\tau \ d\upsilon
\end{align}

\section{Solution to the Basset's equation of motion} \label{appC}
\cite{boggio1907integrazione} successfully integrated the equation of motion \eqref{eq:Basset-motion-equation2}, and obtained an analytical solution to the particle velocity. The method employed by Boggio depends on the Abel integral equation. Let

\begin{equation} \label{eq:Abel-integral0}
    \int_{0}^{t} \frac{1}{\sqrt{t-\tau}} \frac{dV(\tau)}{d\tau} \ d\tau = \phi(t)
\end{equation}

for which a unique solution is

\begin{equation} \label{eq:Abel-integral1}
    \int_{0}^{t} \frac{\phi(\tau)}{\sqrt{t-\tau}} \ d\tau = \pi \big[V(t) - V(0)\big]
\end{equation}

Substituting the definite integral \eqref{eq:Abel-integral0} in \eqref{eq:Basset-motion-equation2}, multiplying by $(t-\tau)^{-1/2}$, integrating from $0$ to $t$, and then eliminating the integral and differentiating with respect to $t$, Boggio reduced the complexity of the problem to the second order differential equation

\begin{equation} \label{eq:Basset-analytical-equation}
    \frac{d^2 V(t)}{dt^2} + \lambda \Big(2 - \frac{\lambda a^2}{\nu} \Big) \frac{dV(t)}{dt} + \lambda^2 V(t) + \frac{h}{\sqrt{t}} - f g \lambda = 0
\end{equation}

where the coefficient $h=\frac{\lambda a}{\sqrt{\pi \nu}} (f g - V_i \lambda)$. \cite{basset1910ondescent} summarised Boggio's work and expressed the solution to the equation of motion \eqref{eq:Basset-analytical-equation} as 

\begin{align} \label{eq:Basset-analytical-solution}
    V(t) = \frac{f g}{\lambda} &+ \Big\{A - h \sqrt{\frac{\pi}{\alpha}} \ \mathrm{erf}(\sqrt{\alpha t}) \Big\} \frac{e^{\alpha t}}{\alpha - \beta} \\
    &+ \Big\{B + h \sqrt{\frac{\pi}{\beta}} \ \mathrm{erf}(\sqrt{\beta t}) \Big\} \frac{e^{\beta t}}{\alpha - \beta}
    \nonumber
\end{align}

where coefficients $A = (fg-\lambda V_i)(1+\beta/\lambda)$ and $B = -(fg-\lambda V_i)(1+\alpha/\lambda)$ are determined by the initial conditions $V(0)=V_i$, and from \eqref{eq:Basset-motion-equation}, $\frac{dV(0)}{dt} = f g - \lambda V_i$. Also, the coefficient $h=\frac{\lambda a}{\sqrt{\pi \nu}} (f g - \lambda V_i)$, and $\alpha,\beta = \frac{\lambda}{2}\{(q-2) \pm \sqrt{q(q-4)}\}$ where $q={\lambda a^2}/{\nu}$. When $V_i=0$, the equation of motion simplifies to

\begin{align} \label{eq:Basset-analytical-solution-zeroini}
    V(t) = \frac{f g}{\lambda} + \frac{f g}{\alpha - \beta} \bigg\{\frac{\lambda a}{\sqrt{\nu \alpha}} \ \mathrm{erfc}(\sqrt{\alpha t}) \ e^{\alpha t} - \frac{\lambda a}{\sqrt{\nu \beta}} \ \mathrm{erfc}(\sqrt{\beta t}) \ e^{\beta t}\bigg\}
\end{align}

This is a solution for the equation of motion \eqref{eq:Basset-motion-equation2} of a small sphere that is initially at rest and falls due to gravity in an infinite fluid that is also initially at rest.

\bibliography{jfm-references}
\bibliographystyle{jfm}
\end{document}